# On-chip quadratically nonlinear photodetector

Yu Zhang[1], Xiaoqing Chen[2,*], Mingwen Zhang[1], Yanyan Zhang[2], Chenyang Zhao[1], Liang Fang[1], Jianlin Zhao[1], Xuetao Gan[1,*]

[1] Key Laboratory of Light Field Manipulation and Information Acquisition, Ministry of Industry and Information Technology, and Shaanxi Key Laboratory of Optical Information Technology, School of Physical Science and Technology, Northwestern Polytechnical University, Xi'an, 710129 China

[2] School of Artificial Intelligence, OPtics and ElectroNics (iOPEN), Northwestern Polytechnical University, Xi'an, 710072, China

E-mail: xiaoqing_chen@nwpu.edu.cn, xuetaogan@nwpu.edu.cn

**Abstract**

Involving deterministically nonlinear photoresponse in on-chip photodetector is intriguing to develop sophisticated functions in photonic integrated circuits, such as in-sensor computing and optoelectronic mixing, though the corresponding devices are still lack of sufficient investigation. Here, we demonstrate an on-chip quadratically nonlinear photodetector (QNPD) by configuring an InSe *p-i-n* homojunction on a silicon waveguide. Telecom-band light guiding in the waveguide couples with the InSe evanescently and is frequency up-converted into visible light via InSe's second-harmonic generation (SHG), which is subsequently absorbed by InSe and finally generates photocurrent under the built-in electric field of the *p-i-n* homojunction. Governed by these sequential processes, the on-chip QNPD presents a quadratic function between photocurrent and optical power. Thanks to the efficient SHG and well-established homojunction in InSe, the QNPD reaches a high normalized responsivity of 37.1 A/W$^2$ and low dark current of 1 pA, representing greatly improved performances among reported nonlinear photodetectors. Benefiting from the extra SHG process, the on-chip QNPD intrinsically incorporates light-light interactions, enabling straightforwardly monitoring all-optically mixing signals electrically. As an example, an array of 16-pixel QNPDs was designed to implement a fully single-shot on-chip autocorrelator without requirement of bulky optics and external cameras, which precisely measures picosecond pulses with high sensitivity of $6.1\times10^{-10}$ $W^2$.

## Introduction

Photonic integrated circuits (PICs) hold great promises to address the massive growth in data traffic of telecom, datacom, computing, sensing, etc.[1-6] To achieve this, chip-integrated photodetectors responsible for optical-to-electrical (O-E) signal conversion in PICs are expected not only to feature with low dark current, high bandwidth, and high responsivity, but also to be qualified for sophisticated optoelectronic functions, such as in-sensor computing, optoelectronic mixing, and temporally mapping ultrashort optical pulses.[7-12] Current chip-integrated photodetectors are primarily based on direct absorption of guiding light to generate photocarriers, which are then converted into photocurrent. Accordingly, the photocurrent is linearly proportional to the optical power, i.e., the photoresponsivity is linear.[7,9,13-17] This attribute makes the on-chip photodetectors difficult to implement the above-mentioned complicated optoelectronic functions due to the absence of light-light interaction or carrier-carrier interaction.

Here, we demonstrate an on-chip photodetector with quadratically nonlinear photoresponse, presenting superior functions over the linear photodetector. Figure 1 schematically displays the structure and operation principle of the proposed chip-integrated quadratically nonlinear photodetector (QNPD). A layered semiconductor Indium Selenide (InSe) is integrated on a silicon photonic crystal (PC) waveguide. The silicon slab is lightly *p*-doped, and the PC waveguide is divided into three electrically separated regions, which could function as individual back-gates of the top InSe layer. By electrostatically gating the InSe layer over separated regions into *p*- or *n*-type, a InSe *p-i-n* homojunction could be configured on the PC waveguide, which provides the built-in electric field to separate the photocarriers in the QNPD.

In the operation of the on-chip QNPD, as illustrated in Figure 1b, the telecom-band light guiding in the silicon PC waveguide couples with the InSe layer via the evanescent field. With this optical excitation, considerable second harmonic generation

(SHG) could be realized beneficial from InSe's strong second-order nonlinearity (*A-process*). Subsequently, the energy-doubled photons of SHG yielded inside the InSe are absorbed in situ (*B-process*) and induce photocarriers, which are finally converted into photocurrent via the *p-i-n* homojunction (*C-process*).

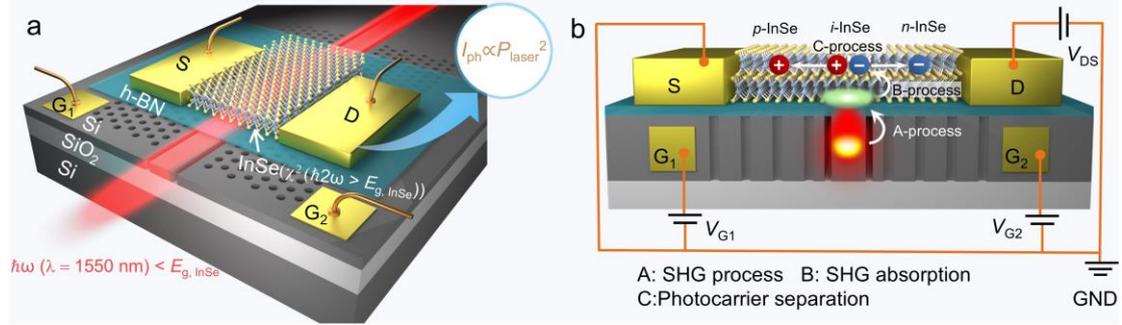

**Figure 1**. Device structure and operation principle of the proposed on-chip QNPD. (a) Three-dimensional schematic of the device, where a InSe layer is integrated on a silicon PC waveguide with a h-BN as the spacing layer. The silicon PC waveguide is designed with three electrically isolated regions by two air-slots, where the two outer PC regions function as back-gates ($G_1$, $G_2$) of the top InSe layer and the central strip waveguide guides light to evanescently couple with InSe. The InSe is electrically contacted with drain (D) and source (S) electrodes. (b) Cross-sectional view of the device illustrating the operation of nonlinear photoresponse. With gate voltages on $G_1$ and $G_2$ with different polarities, the corresponding top InSe regions are electrostatically doped into *p*- or *n*-type, and the InSe region over central strip waveguide remains intrinsic (*i*-type), resulting in *p-i-n* homojunction over InSe channel. Guided light in silicon PC waveguide couples with InSe to yield SHG signal (*A-process*), which is then absorbed by InSe to generate photocarriers (*B-process*) and finally converted into photocurrent via the *p-i-n* homojunction (*C-process*). As a result, the generated photocurrent ($I_{ph}$) follows quadratic dependence on incident power ($P_{laser}$), i.e., $I_{ph} \propto P_{laser}^2$.

The characteristic SHG process in the proposed QNPD expands the capability of silicon chip-integrated photodetectors. First, leveraging the up-conversion of incident photons with energy of $\hbar\omega$ into photons with energy of $\hbar(2\omega)$ by InSe's SHG,[18-19] the integrated QNPD could respond to light with photon energy lower than electronic bandgaps of silicon or InSe. From a fabricated QNPD device, reliable photocurrents were realized in the spectral range of 1500 ~ 1600 nm, which is cutoff by the bandwidth of grating couplers of the PC waveguide. Second, since the SHG photons follow quadratic dependence on incident power,[18,20] the QNPD presents the same trend between the incident optical power ($P_{laser}$) and the generated photocurrent ($I_{ph}$), which can be expressed as $I_{ph} \propto P_{laser}^2$. Third, before the O-E conversion in QNPD, the SHG process has involved the light-light interaction,[21-22] enabling straightforward measurement of all-optically mixing signals electrically. As an example, we design a linear array of QNPDs with 16-pixels along the PC waveguide to measure the temporal overlap of ultrafast optical pulses, which presents a function of single-shot on-chip autocorrelator for accurately measuring pulse width in the level of picoseconds. Assisted by the remarkable SHG, the QNPD-based autocorrelator enables pulse width characterization at peak power as low as 1.75 mW, with a sensitivity of $6.1 \times 10^{-10}$ $W^2$, which is the highest reported among integrated autocorrelators and more than three orders of magnitude higher than commercial two-photon conductivity-based autocorrelators.[12,23-31]

**Results and discussion**

In the device fabrication, the silicon PC waveguide is designed with three electrically isolated regions by two air-slots, as schematically shown in Figure 1. The two outer PC regions with air-hole arrays are employed as two separated back-gate electrodes ($G_1$, $G_2$) for the top integrated InSe layer. The middle strip waveguide region guides light, which evanescently couples with the InSe layer. Before the integration of InSe layer, a thin hexagonal boron nitride (h-BN) layer is transferred on the silicon PC waveguide as the dielectric film. The InSe layer is contacted with drain and source electrodes (D and S). By applying a gate voltage between either left or right PC region and the top InSe layer via the h-BN dielectric, the corresponding InSe region is electrostatically doped into *p*- or *n*-type, depending on the gate voltage polarity. The central strip waveguide region, where no electrical signal is applied, ensures that the corresponding top InSe layer remains intrinsic (*i*-type). This configuration establishes an electrostatically configured *p-i-n* homojunction in the InSe channel. Details of the fabrication process and structural parameters are provided in the Methods section.

Figure 2a presents an optical microscope image of the fabricated on-chip QNPD.

The InSe channel between the source and drain electrodes has a length of approximately 6 μm, and the InSe layer extends 36 μm along the PC waveguide to achieve a long light-InSe interaction. For efficient SHG, multilayer InSe with a thickness of approximately 46 nm is utilized,[20] as confirmed by the atomic-force microscope (AFM; Figure S1, Supporting Information). A 51 nm-thick h-BN dielectric layer ensures a clean van der Waals interface with the InSe channel and effectively prevents unexpected doping by the substrate.[14] Figure S2 in the Supporting Information depicts the transmission spectra of the PC waveguide before and after the InSe integration. The transmission spectral range limited between 1500 ~ 1600 nm is determined by the bandwidth of the grating couplers at the waveguide ends. The integration of InSe layer reduces the waveguide transmission due to the mode scattering at the interface between PC waveguide regions with and without InSe.[14,32] Figure S3 in the Supporting Information displays the photoluminescence (PL) spectrum of the InSe layer with the excitation of a 532 nm laser, revealing its bandgap of approximately 1.26 eV. This value corresponds to a cutoff absorptive wavelength close to 1000 nm, which is consistent with previous reports.[20,33]

To evaluate the electrostatic tunability of the fabricated device, a common gate bias $V_G$ was applied to the two separated silicon back-gates ($V_{G1} = V_{G2} = V_G$) while holding $V_{DS}$ = -3.5 V. As shown in Figure 2b, the InSe channel exhibits clear ambipolar behavior, with *p*-type (hole-dominated) transport at negative $V_G$ and *n*-type (electron-dominated) conduction at positive $V_G$. The drain current $I_{DS}$ varies by more than two orders of magnitude across the sweep, indicating efficient dual-gate control of carrier type and density.

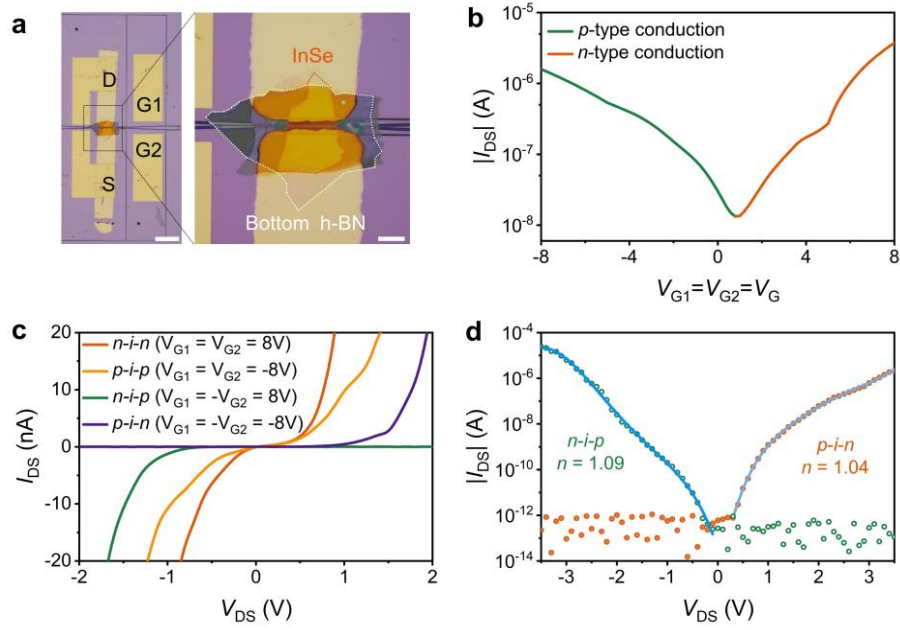

**Figure 2**. Electrical characteristics of the InSe *p-i-n* homojunction constructed on the silicon PC waveguide. (a) Left panel: Optical microscope image of the fabricated device. Scale bar: 100 μm. Right panel: Zoomed image of the active section of the device, consisting of a few-layer InSe flake (red dashed line) and bottom h-BN (white dashed line) dielectric layer. Scale bar: 20 μm. (b) Transfer characteristic ($|I_{DS}|$-$V_G$ curves) measured at $V_{DS}$ = -3.5 V, exhibiting bipolar transport behavior. (c) Output characteristics ($I_{DS}$-$V_{DS}$ curves) under four gate configurations of *n-i-n*, *p-i-p*, *p-i-n* and *n-i-p* homojunctions. (d) Semi-logarithmic plots of $|I_{DS}|$-$V_{DS}$ curves for the *p-i-n* (orange line) and *n-i-p* (green line) configurations, which show a rectification ratio of $10^6$ with an ideality factor of 1.04 and a rectification ratio of $10^7$ with an ideality factor of 1.09, respectively. Both configurations feature an off-state current below 1 pA.

Based on the bipolar property of the InSe device, the output characteristics measured under various gate configurations are presented in Figure 2c. When uniform gate voltages are applied ($V_{G1} = V_{G2} = 8V$ or $V_{G1} = V_{G2} = -8V$), the device acts as an *n-i-n* or *p-i-p* resistor with symmetric $I_{DS}$-$V_{DS}$ curves, consistent with majority-carrier drift-diffusion through an intrinsic section.[34] However, the presence of Schottky barriers results in slight nonlinearity in $I_{DS}$-$V_{DS}$ curve at low bias. Conversely, when opposite gate biases are applied over $G_1$ and $G_2$, as in the *p-i-n* configuration ($V_{G1} = -V_{G2} = 8V$) and the *n-i-p* configuration ($V_{G1} = -V_{G2} = -8V$), a built-in electric field and a depletion region in the channel are established, yielding distinct rectifying diode behaviors.[34-35] At low $V_{DS}$, the currents through the *n-i-n* and *p-i-p* junctions are markedly larger than that through the *p-i-n* and *n-i-p* diodes, suggesting that the *p-i-n* and *n-i-p* junctions dominate the current transport, while the influence of the Schottky barrier remains secondary. To further analyze the rectifying behavior of the *p-i-n* and *n-i-p* junctions, the $I_{DS}$-$V_{DS}$ curves are plotted on a semi-logarithmic scale and fitted with the Shockley diode equation, as shown in Figure 2d. The rapid increase in current under forward bias yields ideality factors *n* of 1.04 and 1.09 for the *p-i-n* and *n-i-p* junctions, respectively, indicating that the current transport is primarily diffusion-driven, with minimal recombination centers or trap states within the junction region.[35-36] The off-state at a reverse bias of |1V| is below 1 pA, with current on/off ratios reaching as high as $10^6$ and $10^7$ for the *p-i-n* and *n-i-p* homojunctions, respectively, further validating the low-power consumption and high quality of these rectifying diodes.

To reveal the fundamental SHG process governing the operation of the on-chip QNPD, we first directly imaged the SHG from the InSe-PC waveguide device under 1550 nm pulsed excitation injected via the input grating coupler. The out-of-plane SHG collected by a CCD forms a continuous emission band strictly coextensive with the InSe-covered waveguide, as depicted in Figure 3a. Its spatial uniformity across input powers from 46 to 92 µW points to a guided-mode interaction sustained over the device length, enabling an effective SHG response accumulated along the propagation path. The effective SHG response results from the combined action of strong guided-mode confinement, which elevates the in-plane field intensity within the InSe layer, and an extended interaction length with substantial mode-InSe overlap across the waveguide footprint.[37] Complementary spectroscopy under 1550 nm excitation exhibits a single, well-defined emission peak at 775 nm with no detectable broadband background, as shown in Figure S4 in the Supporting Information.

The high-efficiency SHG established in the InSe-PC waveguide suggests the feasibility of sensitive nonlinear photodetection. To quantify this, we examined the photoelectric response in the gate-defined InSe *p-i-n* homojunction ($V_{G1}$ = -$V_{G2}$ = 8 V). When a 1550 nm pulsed laser is coupled into the silicon PC waveguide, its photon energy (0.80 eV) lies below the silicon or InSe bandgap and cannot be directly absorbed.

By contrast, the SHG in the InSe layer efficiently up-converts two 1550 nm photons into a 775 nm photon (with energy of 1.60 eV), which is subsequently absorbed in the InSe *p-i-n* homojunction to create electron-hole pairs that are promptly separated by the built-in field (as schematically shown in Figure 1b). Figure 3b displays the $|I_{DS}|$-$V_{DS}$ curves measured in dark and with the illumination of the 1550 nm pulsed laser, the measured $|I_{DS}|$ increases significantly with the increased optical powers, from ~$10^{-12}$A to ~$10^{-6}$A, yielding a light on/off ratio exceeding $10^6$ at $V_{DS}$ = -3 V. This pronounced growth in $|I_{DS}|$ is attributed to the remarkable SHG efficiency within the InSe-PC waveguide, which generates considerable SHG photons that can be effectively converted into current by the high-quality InSe *p-i-n* homojunction.

Enabled by the SHG process, the QNPD exhibits a clear quadratic photoresponse. We varied the pump power and measured the photocurrents. As shown in the log-log plot in Figure 3c (green line), the photocurrent shows a quadratic relationship with optical power, following $I_{ph} \propto P_{laser}^2$ (fit exponent 2.10 ± 0.04). To elucidate the photocurrent mechanism, we simultaneously collected the out-of-plane SHG emission from the InSe channel with a spectrometer. The SHG intensity (orange line in Figure 3c) shows the same quadratic scaling with the pump power (fit exponent 2.09 ± 0.02),[20,38] confirming that the response under 1550 nm excitation arises from SHG-mediated frequency up-conversion in the InSe *p-i-n* homojunction.

Benefiting from the high-efficiency SHG in the waveguide-integrated InSe and the high-quality gate-defined *p-i-n* homojunction, the on-chip QNPD achieves a responsivity of 7.6 mA/W under 1550 nm excitation at an average optical power of 207.2 µW. This represents the highest responsivity reported among nonlinear photodetectors based on two-dimensional materials (Figure 3d),[22,39-47] which corresponds to a normalized responsivity of 37.1 A/W$^2$. The highest responsivity is consistent with the synergy of efficient guided-mode confinement and extended light-InSe interaction length that enhance SHG process, together with low-leakage carrier separation in the high-quality InSe *p-i-n* homojunction.

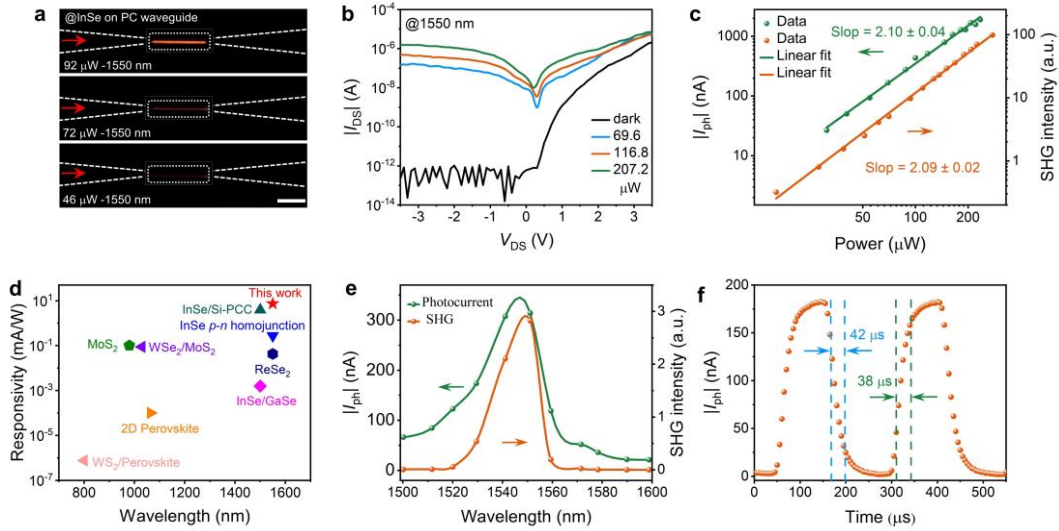

**Figure 3**. SHG and nonlinear photoelectric response of the on-chip QNPD based on InSe *p-i-n* homojunction ($V_{G1}$ = -8V, $V_{G2}$ = 8V). (a) SHG spatial distribution images from the InSe-PC waveguide under excitation by the guided 1550 nm pulsed laser at different optical powers. Scale bar: 20 μm. (b) $|I_{DS}|$-$V_{DS}$ curves of the InSe *p-i-n* homojunction under the excitation of a 1550 nm pulsed laser guided in the waveguide at different optical powers. (c) Laser power dependences of the photocurrent (green line) at $V_{DS}$ = -3.5 V and the SHG intensity (orange line). (d) Comparison of performances of the nonlinear photodetectors based on 2D materials: WS$_2$/Perovskite[39], 2D Perovskite[40], MoS$_2$[41], WSe$_2$/MoS$_2$[42], GaSe/InSe[22], ReSe$_2$[43], InSe *p-n* homojunction[44] and InSe/Si-PCC[45]. (e) Quadratic photocurrent (green line) and SHG intensity (orange line) as a function of wavelength from 1500 to 1600 nm. (f) Time-resolved photoresponse of the QNPD under excitation of modulated 1550 nm pulsed laser.

To further elucidate the photocurrent mechanism, we measured both the photocurrent and the out-of-plane SHG emission from the on-chip QNPD as functions of the incident wavelength, as shown in Figure 3e. By tuning the laser wavelength from 1500 nm to 1600 nm, both the SHG intensity (orange line) and the photocurrent (green line) rise and then fall with a common maximum near 1550 nm. The coincident line shapes indicate that photocarrier generation tracks the waveguide-mediated SHG efficiency. This trend corresponds to the transmission spectrum characteristics of the PC waveguide (see Figure S2 in Supporting Information), indicating that the guided-mode field enhancement governs the SHG strength and, in turn, the photocurrent.

Notably, this SHG-assisted nonlinear detection mechanism enables the InSe-based photodetector to respond sub-bandgap photons by converting them into above-bandgap SHG photons, thereby relaxing the photoresponse threshold from $\hbar\omega > E_g$ to $\hbar\omega > E_g/2$. Consequently, the device achieves an extended spectral response beyond the conventional bandgap limit (~1000 nm) of InSe or silicon, enabling efficient detection across the telecom-band.

To assess the temporal performance of the on-chip QNPD, a mechanical chopper was used to modulate the 1550 nm pulsed laser, and the resulting photocurrent was recorded, as shown in Figure 3f with the InSe channel configured in *p-i-n* configuration. Both the rise and fall times of the quadratically nonlinear photoresponse are approximately 40 μs, measured between 10% and 90% of the maximum photocurrent. This response speed is not affected by the SHG process itself, since SHG is a coherent optical process with an intrinsic femtosecond response and therefore does not limit the response speed.[48] Instead, the measured response speed is governed by electrical factors, primarily the RC time constant and carrier transit in the junction. Higher bandwidth can be achieved by reducing junction capacitance and series resistance, for example by shrinking the overlap region, shortening the carrier path, and optimizing the contacts.[49-50]

Additionally, we extended the study of the quadratic photoelectric responses of the on-chip QNPD to other InSe homojunction configuration with the same device, as detailed in Figure S5 of the Supporting Information. The measurements from the configuration of *n-i-p* homojunction demonstrated results consistent with those obtained from the configuration of *p-i-n* homojunction. Furthermore, similar results were reproduced across different devices, as shown in Figure S6 of the Supporting Information. These measurements indicate that the nonlinear photoresponse of the on-chip QNPD is primarily governed by the SHG in InSe and efficient carrier separation in the gate defined homojunction, rather than by device specific artifacts, thereby establishing the robustness of the platform across junction configurations and samples.

Benefiting from the SHG-mediated nonlinear photoelectric response of the on-chip QNPD, the device intrinsically incorporates light-light interactions, enabling straightforward monitoring of all-optically mixing signals electrically. Leveraging this capability, we design a linear array of QNPDs along the PC waveguide to measure the temporal overlap of ultrafast optical pulses, which presents a function of single-shot on-chip autocorrelator for accurately measuring pulse width. Accurate temporal characterization of ultrafast pulses on PIC platforms remains a key challenge in ultrafast photonics.[51] Photodetectors based on two-photon absorption (TPA) can be implemented in PICs as delayed-scanning autocorrelators, but the intrinsically weak third-order nonlinearity yields low conversion efficiency and limited responsivity.[23-24] Alternatively, several studies image SHG or THG signals in waveguide with external CCDs to achieve single-shot on-chip autocorrelator.[25-26] For example, in our previous work,[25] few-layer InSe integrated on a PC waveguide produced strong SHG at the spatiotemporal overlap of two counter-propagating fundamental pulses, recording the SHG spatial profile with a CCD and mapping it to the time domain-enabled high-precision pulse-width extraction with a sensitivity of $2.4 \times 10^{-8}$ $W^2$. However, reliance on an external CCD hinders true on-chip integration and real-time readout.

To overcome these limitations, based on the on-chip QNPD, we propose a fully integrated, electrically read single-shot on-chip autocorrelator. By constructing an array of QNPD units and using guided-mode SHG as the native light-light interaction channel, the temporal overlap of two fundamentals in the InSe-PC waveguide is converted into above-bandgap SHG photons. At each QNPD sampling site, the InSe QNPD efficiently absorbs these SHG photons and photocurrent is generated with the actuation of the built-in electric field, directly encoding the pulse-overlap distribution as an electrical signal. This design eliminates external optics and enables compact, real-time, on-chip pulse characterization. A schematic of the architecture is shown in Figure 4a.

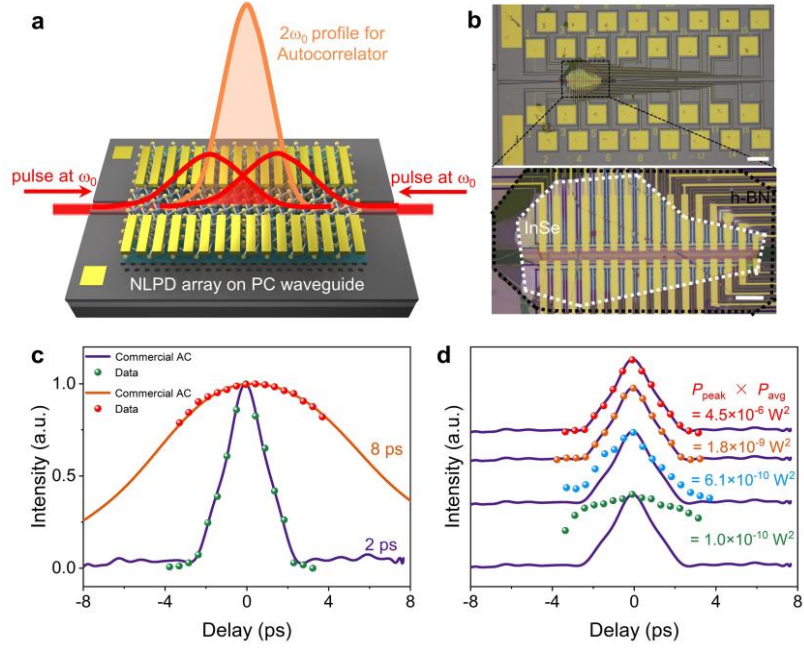

**Figure 4**: Integrated single-shot autocorrelator based on the on-chip QNPD array. (a) Schematic of the QNPD array integrated on a PC waveguide for single-shot on-chip autocorrelator. Two fundamental pulses propagate in opposite directions and overlap within the InSe layer, producing a steady-state SHG spatial profile. The QNPD array detects the SHG signals at specific positions, directly reflecting the temporal overlap of the pulses. (b) Top panel: Optical micrograph of the fabricated on-chip QNPD array used as an integrated autocorrelator. Scale bar: 100 μm. Bottom panel: Enlarged view of the active region of the device, showing 16-pixel QNPD units uniformly spaced at 5 μm intervals along the PC waveguide. Scale bar: 20 μm. (c) Comparison of the autocorrelation results from the on-chip QNPD array and a commercial autocorrelator (AC) for pulse widths of 2 ps and 8 ps at 1550 nm. (d) Autocorrelation results acquired by the fabricated QNPD device (dotted data) for different $P_{peak} \times P_{avg}$ values. The solid line is a result obtained by the commercial autocorrelator at $P_{peak} \times P_{avg} = 1.25 \times 10^{-3}$ W$^2$, which is plotted along with each result obtained from the QNPD device for eye-assistance.

The single-shot on-chip autocorrelator comprises an array of 16-pixel QNPD units uniformly spaced by 5 μm along the PC waveguide. Each pixel features a 6 μm source-drain channel gap, and air slots are inserted between neighboring pixels to ensure electrical isolation and suppress crosstalk. Optical microscope images of the fabricated autocorrelator are shown in Figure 4b. The temporal sampling interval follows $\Delta t = \sqrt{2} \times \Delta z \times n_g / c$,[25-27] where $c$ is the speed of light in vacuum and $n_g$ is the group index of the PC waveguide which is already proven as $n_g$ = 9.5 in our previous study,[25] $\Delta z$ = 5 μm is the pixel pitch (details in Methods). Each QNPD unit operates as a dual gate-defined p-i-n homojunction under $V_{DS}$ = -1.5 V. To evaluate the performance of the QNPD array as a single-shot on-chip autocorrelator, pulse duration measurements were conducted at a wavelength of 1550 nm for pulse widths of 2 ps and 8 ps respectively. The results were compared with those obtained using a commercial autocorrelator with bulky optical components and external CCD camera (APE, NX150). As shown in Figure 4c, for the measurement results of the pulses with different widths, the on-chip QNPD-based autocorrelator achieves excellent agreement with the commercial autocorrelator, confirming the ability of our proposed integrated autocorrelator to accurately characterize different pulse widths.

Sensitivity is a key metric for evaluating an autocorrelator's ability to detect low-power signals, defined as the product of the minimum detectable peak and average powers ($P_{peak} \times P_{avg}$).[23] To quantitatively assess the sensitivity of our integrated autocorrelator, we recorded the pulse width characteristics of the QNPD array under a series of progressively reduced input powers, as shown in Figure 4d. At an average power of 30 μW (corresponding to a peak power of 150 mW and a sensitivity of 4.5 × $10^{-6}$ $W^2$), the integrated autocorrelator obtained an autocorrelation waveform that is highly consistent with the measurement results of a commercial autocorrelator. Even at a significantly low input average power of 0.35 μW (peak power: 1.75 mW; sensitivity: 6.1 × $10^{-10}$ $W^2$), the measured autocorrelation waveform remained clearly resolvable, indicating adequate photocurrent signal quality for pulse width reconstruction. As the power was further reduced to 0.15 μW (peak power: 0.75 mW; sensitivity:1.0 × $10^{-10}$ $W^2$), the trace shows significant broadening and shape deviations due to increasing influence of noise current, marking the signal degradation at this level. Based on these measurements, we confirm that our integrated QNPD-based autocorrelator achieves a sensitivity as high as 6.1 × $10^{-10}$ $W^2$, which is the highest among reported chip-integrated autocorrelators. This sensitivity represents more than two orders of magnitude improvement compared to commercial SHG-based autocorrelators and previously reported on-chip autocorrelators based on TPA detectors (Table 1,

Supporting Information).[12, 23-31] This is attributed to the high nonlinear responsivity of our proposed QNPD and the localized energy density enhancement induced by the interference of counter-propagating pulses within the PC waveguide. These results underline the exceptional performance of our QNPD-based device for ultralow-power, high-sensitivity ultrashort pulse characterization, highlighting its promising potential for large-scale integration within PICs.

**Conclusion**

In conclusion, we have demonstrated an on-chip QNPD, a nonlinear photodetector with quadratic photoresponse, by integrating an InSe *p-i-n* homojunction on a silicon PC waveguide. Relying on strong second-order nonlinearity of InSe, telecom-band light guided in the PC waveguide is frequency up-converted into visible light via InSe's SHG, which is subsequently absorbed by the InSe in situ. The generated photocarriers are then separated by the built-in electric field of the *p-i-n* homojunction, which is electrostatically configured by the back-gate of silicon PC waveguide. Governed by sequential processes in SHG of guiding light, absorption of SHG photons, separation of photocarriers, the on-chip QNPD presents a quadratic function between photocurrent and incident optical power, i.e., $I_{ph} \propto P_{laser}^2$. Thanks to the efficient SHG in InSe layer and well-established homojunction, the QNPD reaches a high normalized responsivity of 37.1 A/W$^2$ and low dark current of 1 pA, representing the highest responsivity and lowest dark current reported among nonlinear photodetectors based on two dimensional materials. The frequency up-conversion involved in the SHG process enable the on-chip QNPD respond to light with photon energy lower than electronic bandgaps of silicon or InSe.

Benefiting from the SHG-mediated nonlinear photoelectric response, the on-chip QNPD intrinsically incorporates light-light interactions, enabling straightforward monitoring of all-optically mixing signals electrically. As an example, a 16-pixel QNPD array on the PC waveguide was fabricated to function as a fully single-shot on-chip autocorrelator. Counter-propagating ultrashort pulses generate a spatial SHG envelope in the InSe layer that is read out electrically at each pixel, which successfully measured optical pulses with different widths of 2 ps and 8 ps with a high sensitivity up to $6.1 \times 10^{-10}$ W$^2$. This architecture eliminates the need for bulky optical components and external CCD cameras, providing a compact, low-power solution for ultrafast pulse width characterization in high-density PICs.

We note that, while a part of reported on-chip photodetectors also could present nonlinear photoresponsivity, for example, *p-n* junction-based photodetectors operating in the saturable mode or avalanche mode, their functions between photocurrent and optical power are varied with respective to the optical powers.[12,52-53] The demonstrated QNPD has a deterministically quadratic function, promises the reliability and signal fidelity. Moreover, besides the autocorrelation measurement of ultrafast laser pulses, the direct photoelectric conversion of all-optically mixing signals in the QNPD simplifies the optoelectronic mixing systems, implying great potentials in on-chip in-sensor computing, optical sampling, etc.[10,22,43]

**Methods**

*Device Fabrication:* The PC waveguides and electrical isolation air-slots were fabricated on a 220 nm thick silicon layer of a silicon-on-insulator (SOI) substrate. A 400-nm thick AR-P 6200.13 resist was exposed by electron-beam lithography (EBL), and the pattern was transferred into silicon by inductively coupled plasma (ICP) etching. The resist was removed with N-methyl-2-pyrrolidone (NMP) followed by a piranha clean; The h-BN and InSe flakes were mechanically exfoliated onto a polydimethylsiloxane (PDMS, Gel-Pak) stamp using the Scotch tape method. Using a precise alignment system, the h-BN and InSe flakes were then sequentially transferred onto the PC waveguide via dry transfer. For the single-device structure in Figure 2a, The Au electrodes with the thickness of 50 nm were prepared by electron beam evaporation (EBE) on a Si substrate using a shadow mask. And then picked up from the substrate with PDMS and transferred onto InSe layer as the source-drain electrodes. For the QNPD array in Figure 4a, the source and drain electrodes were defined by EBL and deposited directly by EBE with 5/50 nm Cr/Au. All transfers were performed in a glove box immediately after exfoliation to minimize interfacial contamination. Devices were annealed at 200 °C for 2 h in Ar with 5% $H_2$ to relieve stress and improve contact quality.

*Photoresponse characterization:* The electrical and optoelectronic measurements were performed at room temperature in ambient conditions using a semiconductor parameter analyzer (PDA FS380 Pro). A picosecond optical parametric oscillator (OPO, 2 ps pulse width, 100 MHz repetition rate, tunable from 1500 nm to 1600 nm) was coupled into a single-mode fiber and aligned to the input grating coupler of the PC waveguide with a six-axis micromanipulator. The transmission optical powers of the PC waveguide were monitored from the other grating coupler by an optical power meter. The semiconductor parameter analyzer collected the nonlinear photocurrent at different optical powers. The power dependence of the SHG was characterized by fixing a single-mode fiber above the emission region to collect the out-of-plane signal, which was then transferred to a spectrometer for spectral analysis. For the single-shot on-chip autocorrelator measurement, splitting optical pulses from the OPO laser at 1550 nm and launching the resulting counter-propagating replicas onto the chip. The two arms of the setup included a polarization controller and a power meter to ensure balanced power on both sides. One arm incorporated a motorized tunable delay line (with a maximum delay of 1500 ps) to match the optical path length on both sides. Note that all of these components could be simplified by using an integrated 3 dB splitter. Each QNPD pixel in the array had a source-drain spacing of 6 μm and an electrode width of 5 μm. The pitch between adjacent QNPDs was 5 μm, forming an array of 16 device pairs with a total length of 160 μm. The time resolution $\Delta t$ of the autocorrelator is given by the relation $\Delta t = \sqrt{2} \times \Delta z \times n_g / c$, where $c$ is the speed of light in vacuum, $\Delta z = 160$ μm is the total array length, and $n_g = 9.5$ is the group index of the PC waveguide. Substituting these values gives a total time delay of approximately 7.16 ps. To assess the accuracy of the integrated autocorrelator, 2 ps and 8 ps pulsed lasers (repetition rate of 40 MHz) were used for the autocorrelation measurements.

**Data availability**

The data that support the findings of this study are available from the corresponding authors upon request.

**Acknowledgements**

This work is supported by National Key R&D Program of China (Grant No. 2022YFA1404800), National Natural Science Foundation of China (Grant Nos. 12374359, 62375225 and 62305270), the Fundamental Research Funds for the Central Universities (Grant No. G2025KY06018).

**Contributions**

Y.Z. conducted the experiments, collected the data, and wrote the manuscript. X.C., X.G. and Y.Z. conceived and designed the experiments. Y.Z. and M.Z. performed the device fabrication and characterization. Y.Y.Z., C.Z., L.F. and X.G. Contributed to the data analysis. X.C., J.Z, and X.G. supervised the project. All authors discussed the results and commented on the manuscript.

**Competing interests**

Competing interests
The authors declare no conflict of interest.